\documentclass[11pt]{article}

\usepackage{a4wide}

\usepackage{amssymb}
\usepackage{eufrak}
\usepackage{epsfig}
\usepackage{graphicx}

\title{
Enhancement of  space-charge induced damping due to reactive impedances for head-tail modes}

\author{
V.~Kornilov,\\
\normalsize{GSI Helmholtzzentrum, Planckstr.\,1,
Darmstadt, Germany},\\
O.~Boine-Frankenheim,\\
\normalsize{GSI Helmholtzzentrum, Planckstr.\,1,
Darmstadt, Germany},\\ 
\normalsize{TU Darmstadt, Schlossgartenstr.\,8, Darmstadt, Germany}}

\begin{document}
\maketitle


\begin{abstract}
Landau damping of head-tail modes in bunches due to spreads
in the tune shift can be a deciding factor for beam stability.
We demonstrate that the coherent tune shifts due to reactive impedances can enhance the space-charge induced damping
and change the stability thresholds (here, a reactive impedance implies the imaginary part of the impedance of both signs).
For example, high damping rates at strong space-charge, or damping of the $k=0$ mode, can be possible.
It is shown and explained, how the negative reactive impedances (causing negative coherent tune shifts similarly to the effect of space-charge) can enhance the Landau damping, while the positive coherent tune shifts have an opposite effect.
It is shown that the damping rate is a function of the coherent mode position
in the incoherent spectrum, in accordance with the concept of the interaction of a collective mode with resonant particles.
We present an analytical model, which allows for quantitative predictions
of damping thresholds for different head-tail modes, for arbitrary space-charge and coherent tune-shift conditions,
as it is verified using particle tracking simulations.
\end{abstract}

\section{INTRODUCTION}

The performance of many present and future hadron ring accelerators
is limited by the effects due to the self-field space-charge.
The transverse head-tail modes in bunches have been initially described
by a theory \cite{sach76, sach72} without space-charge.
A significant progress have been recently achieved
in understanding of the effects of the betatron tune shifts due to space-charge
for the collective eigenmodes in bunches.
Similarly to other beam dynamics issues, the spread in the space-charge tune shift
appeared to be very important.

An "airbag" bunch model, suggested in \cite{blask98}, allowed for an exact
description of the head-tail modes for arbitrary space-charge.
This model does not take into account a spread in the space-charge tune shift which is present in realistic bunch distributions.
Nevertheless, the predictions of the airbag model
appeared to be rather accurate for the eigenfrequencies
and for the eigenmodes in the realistic Gaussian bunches \cite{kornilov-icap09, kornilov_prstab10}.
Experimental confirmations have been demonstrated
in measurements on bunches with space-charge in \cite{kornilov_prstab12, singh2013}.
Theories for realistic bunch distributions have been presented in \cite{burov2009, balb2009},
where an important implication of the space-charge tune spread has been identified.
This is a Landau damping of the coherent modes due to interactions with the incoherent particle spectrum.
In contrast to the case of a coasting beam, the space-charge induced tune spread
provides damping for the bunch eigenmodes.
Damping rates due to the tune spread produced by the longitudinal bunch density variations
have been demonstrated and quantified in \cite{kornilov_prstab10}.

An imaginary impedance of the accelerator facility represents a reactive force and shifts the
coherent frequencies \cite{ng, chao}.
In real accelerators there are many sources of the coherent and incoherent tune shifts.
The detuning wakes can shift the incoherent betatron frequencies \cite{burov99}.
In this work we focus on the driving dipole wakes and on their coherent betatron frequency shifts.
We thus specify $\Delta Q_{\rm coh}$ as the real tune shift of a coherent eigenmode due to a reactive facility impedance.
Typically, the real coherent tune shift is negative, which is also the case for the self-field space-charge tune shift $\Delta Q_{\rm sc}$,
so for the both notations we use the modulus of the negative value.
As the both shifts depress the tunes, it makes the intersections of the coherent frequencies with the
incoherent spectrum possible. This is the main reason for the effect of the reactive impedances
on the space-charge induced damping in bunches which we address in the present paper.
The combination of space-charge with reactive impedances has been discussed in \cite{boine_prstab09, singh2013},
and the enhancement effect of the reactive impedances on damping has been suggested in \cite{kornilov-hb14, kornilov-sc15}.
In this work we present an analytical model and simulations for a systematic description
of the space-charge induced damping with reactive impedances.

In Sec.\,2 we discuss the nature of the space-charge induced damping and how it can depend
on the beam distributions, especially regarding the distribution tails.
The effect of reactive impedances is explained in Sec.\,3, with the analytical model for
arbitrary space-charge conditions and mode index number.
The predictions of the analytical model are analysed for different cases and verified
in Sec.\,4. The work is concluded in Sec.\,5.

\section{SPACE-CHARGE INDUCED DAMPING}

There are two basic sources for the self-field space-charge tune spread in a bunch.
The first one is due to different synchrotron amplitudes of the individual particles, i.e.\ in the longitudinal plane.
The second one is due to different betatron amplitudes, i.e.\ in the transverse plane.
In the both cases, a non-uniform density distribution causes a tune spread.

The line density $\lambda(z)$ variation along the bunch
gives the maximal space-charge tune shift for the given longitudinal position,
\begin{eqnarray}
\Delta Q_{\rm sc} (z) =
\frac{g_\perp \lambda(z) r_p R}{4 \gamma^3 \beta^2 \varepsilon_x} \ ,
\label{fo07}
\end{eqnarray}
where $(2 \pi R)$ is the ring circumference,
$\beta$ and $\gamma$ are the relativistic parameters,
$r_p=q_{\rm ion}^2/(4 \pi \epsilon_0 m c^2)$ is the classical particle radius,
$\varepsilon_x$ is the transverse rms emittance.
This tune shift is the modulus of the negative shift.
The geometry factor $g_\perp$ depends on the transverse distribution,
for the Gaussian profile it is $g_\perp = 2$,
for the K-V beam it is $g_\perp = 1$.
In order to characterize the space-charge force in a bunch, the space-charge parameter is used,
\begin{eqnarray}
q = \frac{\Delta Q_{\rm sc} (0)}{Q_s} \,
\label{fo08}
\end{eqnarray}
which is the tune shift for the rms-eqivalent K-V beam ($g_\perp=1$) in the peak of the line density $(\lambda_0)$,
normalised by the synchrotron tune $Q_s$.
It was shown in \cite{blask98, burov2009, balb2009, kornilov_prstab10, kornilov_prstab12} that this parameter is the key value
to describe the space-charge conditions in a bunch. This is especially true for the analytical models, as it will be also shown below. The usage of the rms-eqivalent K-V beam for the comparisons between different beam distributions has been proven to be adequate, the same is true for comparisons with the experiments. 
During the further discussion we use the notation $\Delta Q_{\rm sc}$ for $\Delta Q_{\rm sc} = q Q_s$ from Eq.\,(\ref{fo08}).

Another reason for the space-charge tune spread is the different amplitudes of the
transverse betatron oscillation of the individual particles.
The particles with small amplitudes have the tune shift close to the local maximal space-charge tune shift from Eq.\,(\ref{fo07}).
The particles with large amplitudes have smaller tune shifts, technically down to zero,
depending on the transverse distribution and on its truncation.
A beam with the transverse K-V distribution has no space-charge tune spread of this kind.

\begin{figure}[!h]
\centering
\includegraphics[width=0.46\linewidth]{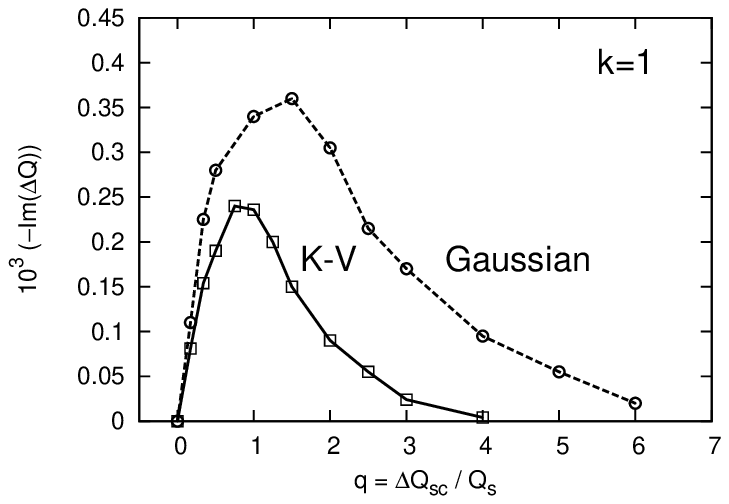}
\includegraphics[width=0.46\linewidth]{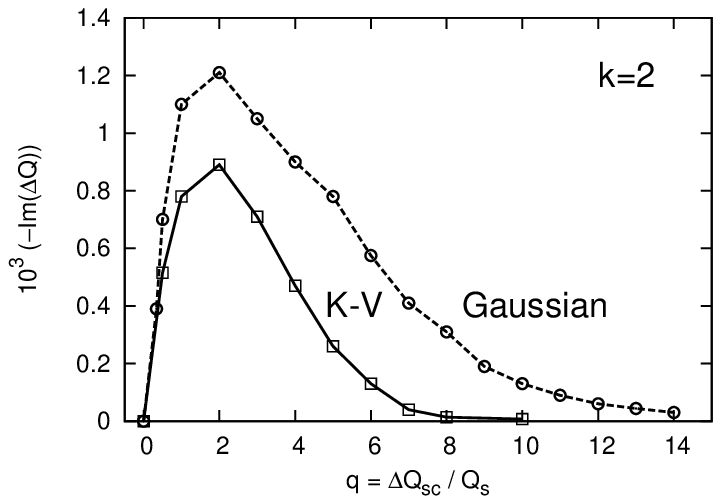}
\caption{\label{fg07}
Space-charge induced damping rates from particle
tracking simulations for $k=1$ and $k=2$ head-tail modes.
Solid lines correspond to the  longitudinal Gaussian and transverse K-V bunches,
as presented in \cite{kornilov_prstab10}.
Dashed lines are for the Gaussian (longitudinal and transverse) bunches.
}
\end{figure}

Damping of the head-tail modes exclusively due to the space-charge tune spread
from the longitudinal plane has been studied in \cite{kornilov_prstab10}.
There, a K-V distribution has been assumed for the transverse plane.
A good agreement with the results from \cite{kornilov_prstab10} has been recently reported in \cite{macridin2015},
in simulations with a different code and different methods.
One might expect that an additional tune spread due to
the transverse profile should enhance damping.
This is confirmed in our particle tracking simulations presented in Fig.\,\ref{fg07},
where the space-charge induced damping for the Gaussian bunch in the longitudinal and transverse planes
is compared with the results from \cite{kornilov_prstab10}.
Additional space-charge tune spread provides higher damping rates and larger $q-$regions for a chosen damping level.

For the particle tracking simulations we use the code PATRIC \cite{boine2006},
similarly to the simulations presented in \cite{kornilov_prstab10}.
The code has been validated using exact analytical prediction \cite{kornilov-icap09},
for the space-charge effects and for the coherent effects.
For the transverse space-charge force, a frozen slice-to-slice electric field model is used,
i.e.\ a fixed potential configuration which follows
the center of mass for each bunch slice.
This approach is justified in the rigid-slice regime \cite{burov-lebed2009, burov2009}.
Thus, it is physically reasonable for moderate and strong space charge.
The recent study \cite{kornilov-sc15} suggests to confirm the applicability of the frozen space-charge model.
The method to excite the head-tail eigenmodes and to determine the damping decrements
is described in \cite{kornilov_prstab10}.
The simulations for Fig.\,\ref{fg07} have been performed for the chromaticity equal zero.
The chromaticity does not affect damping, which is also clear in the formalism of \cite{burov2009},
and has been verified in our simulations.

\section{ANALYTICAL MODEL}

The basic mechanism of Landau damping is the interaction between a collective wave and the resonant particles.
In the case of the head-tail modes in bunches these roles are played by a coherent bunch eigenmode and the individual particle betatron oscillations.
The damping rate is then proportional to the number of the resonant particles. It can be especially clear in the simulations
for different longitudinal and transverse distributions in a bunch. Allowing more particles in the distribution tails,
both longitudinal and transverse, provides stronger damping.
In the experimental conditions, the bunch distributions can be
very complicated, the tail truncations depend
on the transverse acceptance and on the momentum acceptance of the machine.
It should be however possible to find general findings for different distributions and tail truncations. In our simulations
we consider the example with the Gaussian distribution, and with the truncations at 2.5 of the standard deviation.

A certain damping rate can be reached if the coherent oscillation
interacts with a large enough number of resonant particles.
The particle incoherent spectrum is provided by tune shifts of different nature.
In the case of space-charge in a Gaussian bunch,
this distribution is non-zero between $(- 2 q Q_s)<\Delta Q<0$.
The particles with tune shifts just below zero ($\Delta Q < 0$), are in the distribution tails, which correspond to particles with large amplitudes.
A growing beam intensity with a fixed beam distribution means a growing space-charge tune shift.
If we consider resonant particles with the fixed oscillation amplitudes, longitudinal and transverse,
we see that the incoherent shift of these particles increases linearly with the intensity parameter,
\begin{eqnarray}
\Delta Q_{\eta} = - \eta  \Delta Q_{\rm sc} .
\label{fo02}
\end{eqnarray}
Here, the parameter $\eta$ is introduced in order to relate the coherent lines to the incoherent spectrum in the upcoming discussion, and to indicate the fact that space-charge tune shifts grow linearly with the beam intensity (for a fixed emittance).
The value of $\eta$ in a specific bunch depends on the chosen damping rate and on the particle
distribution in the bunch.

The coherent tune shift due to space-charge can be analytically estimated
using the airbag model \cite{blask98},
\begin{eqnarray}
\frac{\Delta Q_k}{Q_s} = - \frac{q}{2} \pm \sqrt{\frac{q^2}{4} + k^2} ,
\label{fo09}
\end{eqnarray}
where "+" is for modes $k \geq 0$. As we discuss in the Introduction, this expression
provides a rather accurate prediction for the eigenfrequencies even in the Gaussian bunches, especially for strong space-charge.
This has been confirmed in simulations, experiments, and theoretical studies \cite{burov2009, balb2009, kornilov_prstab10, kornilov_prstab12}.

The head-tail modes drop in frequency for stronger space-charge.
Thus the coherent mode can cross the border of the damping with a chosen damping rate.
Since the head-tail modes with $k>0$ are close to the upper tail of the incoherent spectrum (close to the bare tune $Q_0$),
there is a top border for the chosen damping.
If the coherent mode has a frequency below this border the damping rate will exceed the border decrement.
The conditions for a chosen damping rate should be estimated by the relation of the border Eq.\,(\ref{fo02})
to the coherent $k-$mode frequency which is modulated by the
harmonics of the synchrotron frequency \cite{balb2009, kornilov_prstab10},
\begin{eqnarray}
\Delta Q_{k} - k Q_s,
\label{fo14}
\end{eqnarray}
where $\Delta Q_k = Q_k - Q_0$ is the coherent frequency of the mode $k$.
The coherent line shifts are positive ($\Delta Q_k > 0$) for $k>0$, and the incoherent tune shifts due to space-charge are negative.
This damping model thus also explains why damping is still possible.

\begin{figure}[!h]
\centering
\includegraphics[width=0.5\linewidth]{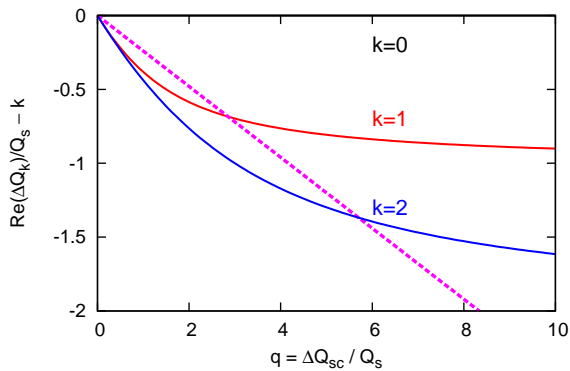}
\caption{\label{fg01}
Positioning of the head-tail modes from Eqs.\,(\ref{fo09}, \ref{fo14}) with respect to the incoherent spectrum border
for the damping analysis.
The dashed line indicates the border Eq.\,(\ref{fo02}) of the chosen damping $\eta = 0.24$.
}
\end{figure}

Figure\,\ref{fg01} illustrates the above discussion and shows the coherent head-tail frequencies modulated by the
harmonics of the synchrotron frequency for the modes $k=1$ and $k=2$.
From the simulations with a Gaussian bunch, see Fig.\,\ref{fg07},
we have determined $\eta = 0.24$, which is shown by the dashed line in Fig.\,\ref{fg01}.

There is a value of the $q-$parameter where the mode crosses the damping border.
It can be predicted using the expression
for the coherent head-tail eigenfrequency Eq.\,(\ref{fo09}).
Comparing $(\Delta Q_k - k Q_s)$ with the damping border Eq.\,(\ref{fo02}), we come to the conclusion that a mode crosses the border at $q=0$, and at
\begin{eqnarray}
q = k \frac{1-2 \eta}{\eta (1-\eta)} \ .
\label{fo10}
\end{eqnarray}
This gives a prediction for the range of the space-charge parameter $q$ for
a chosen damping rate for arbitrary mode index.
Figure\,\ref{fg01} illustrates this: for very small $q$ each head-tail mode
enters the incoherent particle spectrum and thus experiences a strong damping (see Fig.\,\ref{fg07}).
As the coherent line crosses the damping border again,
the damping becomes weaker due to a smaller number of the resonant particles.
There is a smooth transition between damping regimes as the number of resonant particles changes.
Estimations for $q$ where the modes cross the damping borders
Eq.\,(\ref{fo10}) provide $q \approx$ 2.8 for $k=1$ and $q \approx$ 5.7 for $k=2$ (here $\eta=0.24$),
as we can also observe in Fig.\,\ref{fg01}.
All these observations are supported by particle tracking results for Gaussian bunches in Fig.\,\ref{fg07}.

The damping rates decrease at strong space-charge ($q>k$). For Gaussian bunches,
the scaling Im$(\Delta Q ) \sim - 1 / q^3$ has been predicted in \cite{burov99}. The drop of the damping rate for higher $q$
is in agreement with the concept of the coherent frequency shift Eq.\,(\ref{fo14}) towards lower density in the incoherent spectrum.
The fast decrease of the damping rate reflects the fact that the coherent lines are shifted further into the tails of the incoherent distribution .
For realistic bunches, the specific scaling depends on the bunch distributions and the tail truncations.
Below we should demonstrate that the effect
of reactive impedances can lead to high damping rates also for strong space-charge.

\section{DAMPING WITH REACTIVE IMPEDANCES: ANALYTICAL MODEL}

\begin{figure}[!h]
\centering
\includegraphics[width=0.5\linewidth]{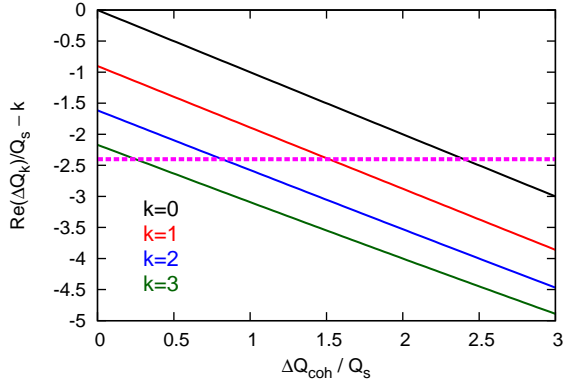}
\caption{\label{fg06}
Combined effect of a reactive impedances and space-charge
on the head-tail mode positioning according to Eq.\,(\ref{fo01}) for $q=10$.
The dashed line indicates the damping border $\Delta Q_\eta$.
}
\end{figure}

The additional effect of a coherent tune shift has been included into the airbag theory in \cite{boine_prstab09},
\begin{eqnarray}
\Delta Q_k = -\frac{\Delta Q_{\rm sc}+\Delta Q_{\rm coh}}{2} \pm
\sqrt{\frac{(\Delta Q_{\rm sc}-\Delta Q_{\rm coh})^2}{4} + k^2 Q_s^2} ,
\label{fo01}
\end{eqnarray}
where "+" is for modes $k \geq 0$.
The coherent tune shift $\Delta Q_{\rm coh}$ of the head-tail mode $k$ is the result of the interaction
with the imaginary part of the facility impedance. It can be calculated as an integral over the bunch spectrum
with the frequency-dependent reactive impedance \cite{sach76, ng, chao}. For a broadband-type impedance
(with respect to the bunch spectrum),
the neighboring modes have very close $\Delta Q_{\rm coh}$, for example for the image charges.
Figure\,\ref{fg06} shows the combined effect of space-charge with the coherent tune shift in this case.
For a narrowband impedance (with respect to the bunch spectrum), the coherent tune shifts can be very different and can contribute to the
excitation of the transverse mode coupling instability \cite{blask98, burov2009, chao} which is not in the scope of this paper.

The difference between the effect of an external impedance and space-charge on head-tail modes is
especially obvious for the $k=0$ mode. Space-charge alone does not affect the $k=0$ mode,
$\Delta Q_{k=0} = 0$. In contrast, a facility impedance alone shifts the $k=0$ mode,
$\Delta Q_{k=0} = - \Delta Q_{\rm coh}$.

There is an important implication from Eq.\,(\ref{fo01}).
For increasing $\Delta Q_{\rm coh}$ (and a fixed $\Delta Q_{\rm sc}$),
each head-tail mode crosses the damping border
and enters the regime with a stronger damping, see Fig.\,\ref{fg06}.
Some of the modes are damped shortly above $\Delta Q_{\rm coh}=0$, for example $k=3$ in Fig.\,\ref{fg06}.
Even the mode $k=0$ is affected by the space-charge induced damping after certain $\Delta Q_{\rm coh}$.
This implies that the coherent effects can change strongly the damping conditions. High damping rates
at strong space-charge, damping of the $k=0$ mode, etc., are possible with reactive impedances,
in contrast to the previously discussed \cite{burov2009, kornilov_prstab10} case of the self-field space-charge only.

The coherent tune shift at which a head-tail mode Eq.\,(\ref{fo01}) crosses the damping border
can be determined using Eq.\,(\ref{fo02}).
The corresponding solution is
\begin{eqnarray}
\Delta Q_{\rm coh} = \Delta Q_{\rm sc} \eta - k Q_s \frac{1-\eta}{k / q + 1-\eta} .
\label{fo05}
\end{eqnarray}

For the $k=0$ mode, this expression is
\begin{eqnarray}
\Delta Q_{\rm coh} = \Delta Q_{\rm sc} \eta \ .
\label{fo04}
\end{eqnarray}

\begin{figure}[!h]
\centering
\includegraphics[width=0.5\linewidth]{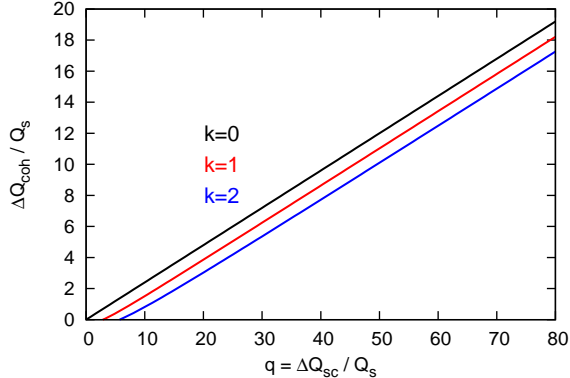}
\caption{\label{fg03}
Coherent tune shift $\Delta Q_{\rm coh}$ needed for a chosen damping rate
for different head-tail modes and for different strength of space-charge in bunches
as predicted by Eq.\,(\ref{fo05}).
}
\end{figure}

For strong space-charge $q \gg k$ Eq.\,(\ref{fo05}) takes a simpler form,
\begin{eqnarray}
\Delta Q_{\rm coh} = \Delta Q_{\rm sc} \eta - k Q_s .
\label{fo06}
\end{eqnarray}
This means that the damping border grows linearly with $\Delta Q_{\rm sc}$ (or $q$).
For the higher-order head-tail modes, the necessary $\Delta Q_{\rm coh}$ is equidistantly shifted by $k Q_s$.
Figure\,\ref{fg03} demonstrates that the coherent tune shift
needed for the chosen damping rate can be easily predicted for arbitrary space-charge conditions.

\section{COMPARISONS BETWEEN THEORY AND SIMULATIONS}

\begin{figure}[!h]
\centering
\includegraphics[width=0.46\linewidth]{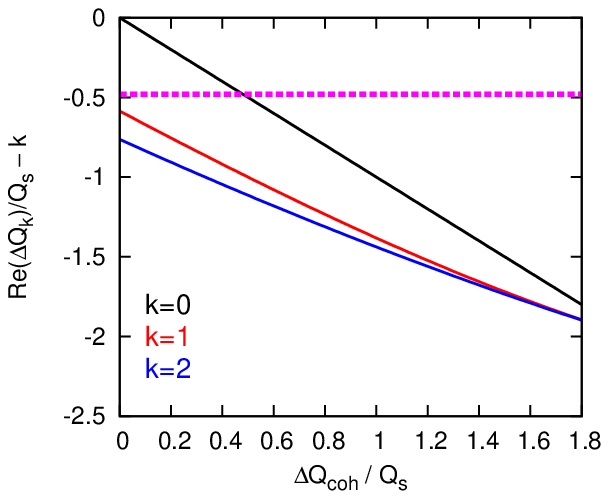}
\includegraphics[width=0.46\linewidth]{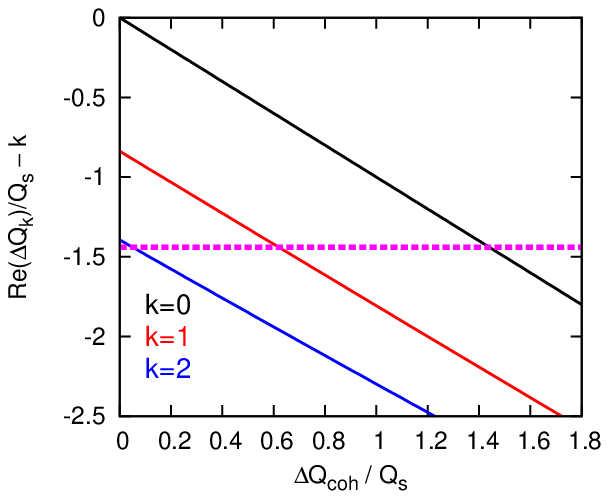}
\caption{\label{fg02}
Effect of the reactive impedances on damping for different space-charge tune shifts according
to Eqs.\,(\ref{fo02}, \ref{fo01}).
Left plot: $q = \Delta Q_{\rm sc}/Q_s=2$, right plot: $q=6$.
The dashed lines indicate the damping border.
}
\end{figure}

\begin{figure}[!h]
\centering
\includegraphics[width=0.46\linewidth]{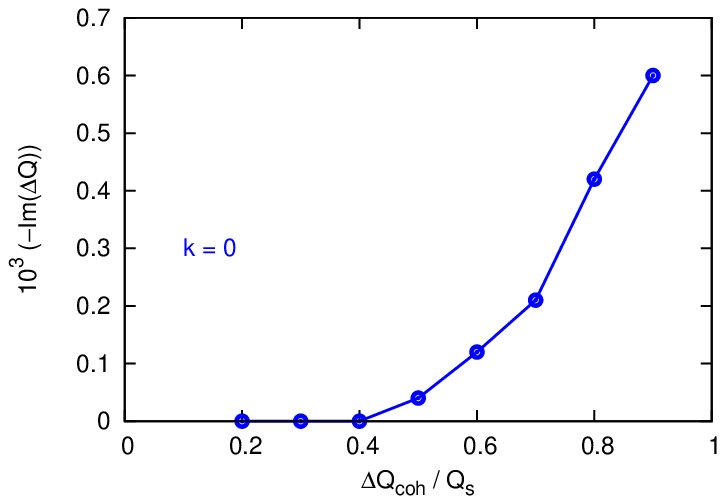}
\includegraphics[width=0.46\linewidth]{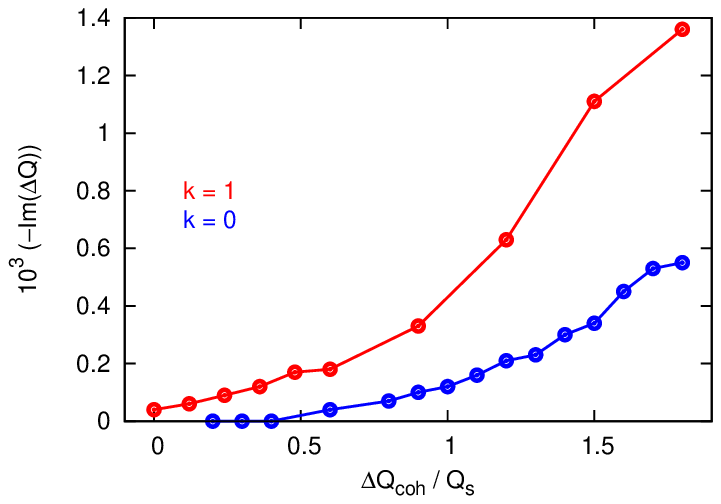}
\caption{\label{fg04}
Effect of the reactive impedances on damping obtained from the particle
tracking simulations.
Left plot: $q = \Delta Q_{\rm sc}/Q_s=2$, right plot: $q=6$.
Note that without reactive impedances the $k=0$ mode experiences no effect
of space charge and no damping.
}
\end{figure}

We verify the basic predictions of the analytical model Eq.\,(\ref{fo05})
using particle tracking simulations.
Figure\,\ref{fg02} shows the positioning of the head-tail modes
with respect to the damping border for $\eta=0.24$. The damping conditions
are compared for the moderate space-charge $q=2$ (left plot)
and for the stronger space-charge $q=6$ (right plot).
This discussion corresponds to a beam with a fixed intensity,
but with an increasing transverse imaginary impedance.
The mode $k=2$ is always damped in the both plots of Fig.\,\ref{fg02}.
The mode $k=1$ is always damped for $q=2$,
but has a region outside the strong damping for $q=6$.
The mode $k=0$ experiences no damping without coherent tune shifts,
but it is affected by damping for increasing $\Delta Q_{\rm coh}$.
For $q=6$, the border of the chosen damping is higher in $\Delta Q_{\rm coh}$
than for $q=2$.
The mode $k=1$ for $q=6$ has only a weak damping without coherent effect,
and around $\Delta Q_{\rm coh}=0.5 Q_s$ the damping becomes stronger.
Results of the simulations in Fig.\,\ref{fg04} confirm these observations.

\begin{figure}[!h]
\centering
\includegraphics[width=0.5\linewidth]{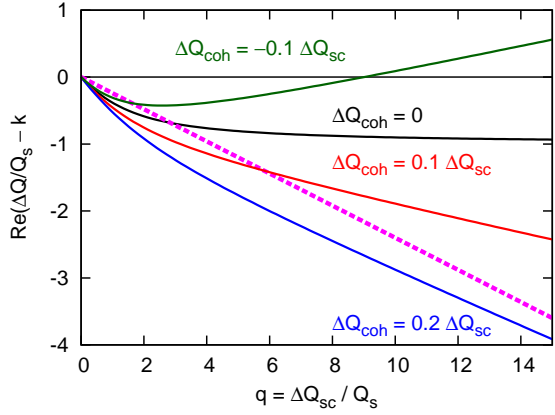}
\caption{\label{fg05}
Effect of the reactive impedances on the $k=1$ head-tail mode according to Eq.\,(\ref{fo01})
for different coherent tune shift
in the case of a fixed $\alpha = \Delta Q_{\rm coh} / \Delta Q_{\rm sc}$.
The dashed line indicates the damping border Eq.\,(\ref{fo02}) for $\eta = 0.24$.
}
\end{figure}

\begin{figure}[!h]
\centering
\includegraphics[width=0.5\linewidth]{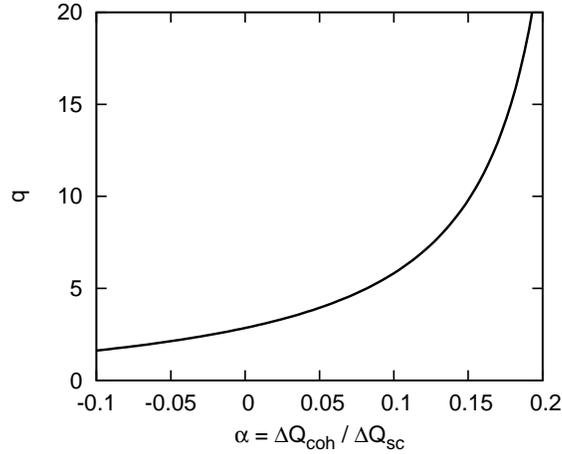}
\caption{\label{fg09}
The space-charge parameter at which the head-tail mode $k=1$
crosses the damping border as given by Eq.\,(\ref{fo11})
with $\eta=0.24$.
}
\end{figure}

In the next case we discuss
an increasing $\Delta Q_{\rm sc}$ with a fixed $\Delta Q_{\rm coh} / \Delta Q_{\rm sc}-$ratio.
This corresponds
to beams of an increasing intensity and with a constant transverse emittance.
The imaginary impedance is also fixed, for example, the vacuum pipe.
Figure\,\ref{fg05} shows the prediction of the analytical model for the mode $k=1$
with the different values of the parameter $\alpha = \Delta Q_{\rm coh} / \Delta Q_{\rm sc}$.
The mode crosses the damping boundary at higher $q-$values
for larger $\alpha-$ratios.
The condition for this crossing for arbitrary $k$ can also be obtained from Eqs.\,(\ref{fo02},\,\ref{fo01}),
\begin{eqnarray}
q = k \frac{\alpha + 1 - 2 \eta}{(\eta - \alpha) (1-\eta)} .
\label{fo11}
\end{eqnarray}
This applies for $\alpha<\eta$ and is plotted in Fig.\,\ref{fg09} for $\eta=0.24$.
The $q-$value of the transfer across the damping border
goes to infinity as $\alpha$ approaches $\eta$.

\begin{figure}[!h]
\centering
\includegraphics[width=0.5\linewidth]{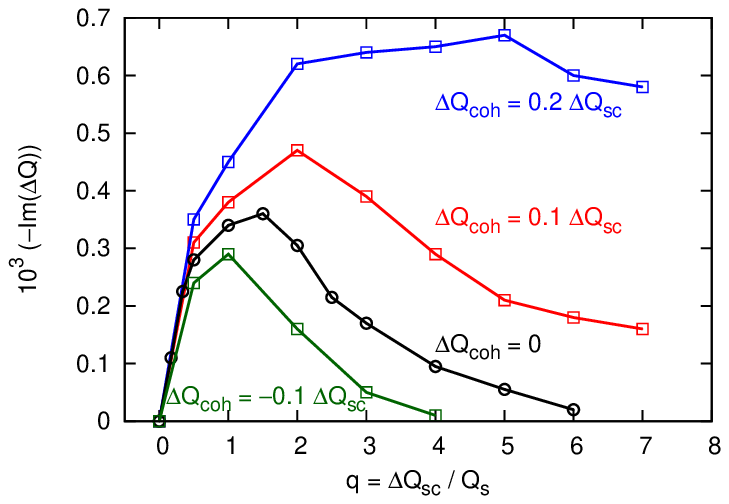}
\caption{\label{fg08}
Damping decrement of the $k=1$ head-tail mode obtained from the particle tracking simulations
for a Gaussian bunch, $Q_s=0.01$,
for different coherent tune shift due to reactive impedances.
The chosen parameters correspond to the analytical model predictions in Fig.\,\ref{fg05}.
}
\end{figure}

Predictions of the analytical model in Figs.\,\ref{fg05},\,\ref{fg09}
are compared with the simulation results in Fig.\,\ref{fg08}.
For the two cases $\Delta Q_{\rm coh} = 0$ and $\Delta Q_{\rm coh} = 0.1 \Delta Q_{\rm sc}$,
there is a region with the strong damping which transfers to a weaker damping by higher $q$.
In the case of $\Delta Q_{\rm coh} = 0.1 \Delta Q_{\rm sc}$
the damping is stronger and has a larger $q-$region.
Within the considered parameters, the mode stays in the strong damping
for $\Delta Q_{\rm coh} = 0.2 \Delta Q_{\rm sc}$.
Correspondingly, the coherent line remains below the damping border in Fig.\,\ref{fg05}.
Finally, a negative coherent shift $\Delta Q_{\rm coh}$ pushes the head-tail mode out of the
damping region and makes the damping weaker, and the damping region smaller.
This can be seen for the case with $\Delta Q_{\rm coh} = - 0.1 \Delta Q_{\rm sc}$ in the analytical model Figs.\,\ref{fg05},\,\ref{fg09},
and it is in a good agreement with the simulations in Fig.\,\ref{fg08}.
Figure\,\ref{fg08} also illustrates the enhancement of Landau damping due to an external impedance, especially for stronger space-charge.

\begin{figure}[!h]
\centering
\includegraphics[width=0.5\linewidth]{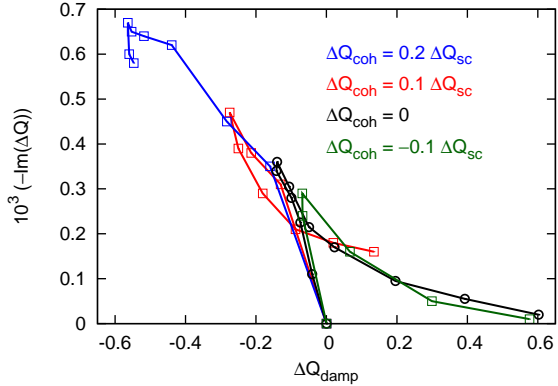}
\caption{\label{fg10}
Damping rates from Fig.\,\ref{fg08} vs.\ the mode-spectrum distance Eq.\,(\ref{fo12}).
The plot demonstrates that the space-charge induced damping depends directly
on the coherent mode position
with respect to the incoherent spectrum, irrespective the space-charge and $\Delta Q_{\rm coh}$ conditions.
}
\end{figure}

Using Fig.\,\ref{fg08} we can test
the suggestion that the space-charge induced damping is directly related to the position
of the coherent line with respect to the damping border in the incoherent spectrum.
For every $q-$value and $\Delta Q_{\rm coh}$ the distance between the modulated
theoretical coherent frequency Eq.\,(\ref{fo01}) and the damping border Eq.\,(\ref{fo02}),
\begin{eqnarray}
\Delta Q_{\rm damp} = \Delta Q_k - k Q_s - \Delta Q_\eta ,
\label{fo12}
\end{eqnarray}
can be calculated and plotted as the horizontal axis, see Fig.\,\ref{fg10}.
The values $\Delta Q_{\rm damp}<0$ imply the position of the mode beyond the spectrum part $\eta$ such that
the resulting damping rate should be higher than that at the damping border.
The points at $q=0$ in Fig.\,\ref{fg08} correspond to $\Delta Q_{\rm damp}=0$ in Fig.\,\ref{fg10}.
As we observe from Fig.\,\ref{fg05}, the coherent lines dive below the damping border
for small $q$, which corresponds to $\Delta Q_{\rm damp}<0$ in Fig.\,\ref{fg10}.
With increasing $q$, the modes come back to the damping border,
and, in fact, the decrements return on the close path back to $\Delta Q_{\rm damp}=0$ (see Fig.\,\ref{fg10}).
The exception is $\Delta Q_{\rm coh}=0.2 \Delta Q_{\rm sc}$ (blue), which stays well below the damping border,
its $\Delta Q_{\rm damp}$ remain strongly negative and the decrements stay high.
For the cases $\Delta Q_{\rm coh}=0$ (black) and $\Delta Q_{\rm coh}=-0.1 \Delta Q_{\rm sc}$ (green),
the modes gain large $\Delta Q_{\rm damp}>0$ and the damping becomes weak.
All these damping rates are located fairly close to a universal line of the space-charge induced damping.
This demonstrates that the damping rate is a monotonic function of the coherent mode position
in the incoherent spectrum, irrespective the space-charge and $\Delta Q_{\rm coh}$ conditions.

\section{CONCLUSIONS}

Space-charge induced Landau damping in bunches has been studied
using an analytical model and particle tracking simulations.
It has been demonstrated that two sources of the space-charge tune spread, longitudinal and transverse, contribute to
the resulting damping.
A specific damping rate (for example, strong enough damping to stabilise an instability) can be considered as the damping border.
The parameter $\eta$ for the shift of the coherent mode into the incoherent spectrum Eq.\,(\ref{fo02})
can then be defined, depending on the distribution in bunch tails and on the chosen damping border.

A model for the space-charge induced damping in bunches is presented.
Using the airbag-based analytical model \cite{blask98} for the head-tail model
with arbitrary space-charge Eq.\,(\ref{fo09}), and the border of the incoherent spectrum
for a chosen damping Eq.\,(\ref{fo02}), the $q-$region of the space-charge induced damping
can be predicted by Eq.\,(\ref{fo10}).
This result is confirmed by the particle tracking simulations.
Enhancement of Landau damping is demonstrated for all considered head-tail modes.

With the help of the analytical expression [Eq.\,(\ref{fo01})] from \cite{boine_prstab09},
a model for the damping border has been suggested, Eq.\,(\ref{fo05}).
For the mode $k=0$ and for strong space-charge, there are simple relations Eqs.\,(\ref{fo04},\,\ref{fo06}).
This model gives a universal prediction for the coherent tune shift needed for the damping, Fig.\,\ref{fg03}, for arbitrary
space-charge conditions and for different head-tail modes $k$.

It has been discussed \cite{burov2009, balb2009, kornilov_prstab10, kornilov_prstab12}
that in the case of self-field space-charge only the mode $k=0$
can not be affected by space-charge and by related damping.
Here we have demonstrated that the analytical model (Fig.\,\ref{fg02}) and the simulations (Fig.\,\ref{fg04}) 
consistently show the conditions where the mode $k=0$ can be damped.

Effects of reactive impedances on different modes $k$ are discussed for
a fixed $q = \Delta Q_{\rm sc} / Q_s$ (constant beam intensity and growing reactive impedance)
and for a fixed $\alpha=\Delta Q_{\rm coh} / \Delta Q_{\rm sc}$
(constant reactive impedance and growing beam intensity).
The both cases reveal different interesting properties in the space-charge induced damping with reactive impedances.
The analytical model Eqs.\,(\ref{fo10},\,\ref{fo05},\,\ref{fo11}) provide
specific predictions for the transition across the damping border for different modes $k$,
which are confirmed by the particle tracking simulations.
The analysis of the simulation results has also demonstrated (see Fig.\,\ref{fg10}) that the space-charge induced damping depends directly
on the coherent mode position
with respect to the incoherent spectrum, irrespective the space-charge and the reactive impedance conditions.

It has been shown that the coherent tune shifts
can have a strong effect on the damping due to space-charge in bunches.
Landau damping can be dramatically enhanced due to the effect of reactive impedances.
Thus the reactive impedances and
other $\Delta Q_{\rm coh}-$sources should be taken into account in an instability threshold analysis.
On the other hand, the space-charge induced damping in bunches can be enhanced
by employing additional reactive impedances, in a passive (dedicated hardware elements)
or in an active (a reactive feedback \cite{ng, ruth83}) manner.

\end{document}